# On the nature of the reentrant effect in susceptibility of mesoscopic cylindrical samples

G.A. Gogadze

*B. Verkin Institute for Low Temperature Physics and Engineering of NAS of Ukraine,*

*47, Lenin Av., 61103 Kharkov,  Ukraine, e-mail: gogadze@ilt.kharkov.ua*

Theory of the reentrant effect in susceptibility of mesoscopic cylindrical *NS* samples is proposed, which is essentially based on the properties of the Andreev levels. The specific feature of the quantum levels of the structure is that in a varying magnetic field (or temperature) each level periodically comes into coincidence with the chemical potential of the metal. As a result, the state of the system becomes strongly degenerate and the amplitude of the paramagnetic contribution to the susceptibility increases sharply.

In 1990 Mota and co-workers [1] detected a surprising behavior of the magnetic susceptibility of a cylindrical *NS* structure (*N* and *S* are for the normal metal and superconductor, respectively) at very low temperature ($T < 100$ *mK*). The external magnetic field was applied parallel to the *NS* boundary. It was most intriguing that a decrease in the sample temperature below a certain point $T_r$ (in a fixed field) produced the reentrant effect: the decreasing magnetic susceptibility of the structure unexpectedly started growing. A similar behavior was observed with the isothermal reentrant effect in a field decreasing to a certain value $H_r$ below which the susceptibility started to grow sharply. The sample were superconducting *Nb* wires with a radius *R* of tens of *μm* coated with a thin layer *d* of very pure *Ag*.  It is emphasized in [2] that the detected magnetic response of the *NS* structure is similar to the properties of the persistent currents in mesoscopic normal rings. It is assumed [1 − 5] that the reentrant effect reflects the behavior of the total susceptibility χ of the *NS* structure: the paramagnetic contribution is superimposed on the Meissner effect-related diamagnetic contribution and nearly compensates it. The anomalous behavior of susceptibility was also observed in *AgTa*, *CuNb* and *AuNb* structures [2,4].

The origin of paramagnetic currents in *NS* structures was discussed in a number of theoretical studies. Bruder and Imry [6] analyzed the paramagnetic contribution to susceptibility taking into account the paths of the quasi-particles that do not collide with the superconducting boundary. The authors note an appreciable paramagnetic effect in the physical situation under discussion. However, the ratio derived by them for the paramagnetic and diamagnetic contributions is rather small, and cannot account for the experimental results [1 −−5].

Fauchere, Belzig and Blatter [7] explain the high paramagnetic effect assuming the pure repulsive electron-electron interaction in noble metals. The proximity effect in the normal metal induces the order parameter whose phase is π-shifted against the $\Delta_s$-phase of the superconductor. This leads to paramagnetic instability of the Andreev states and the density of states of the *NS* structure exhibits a peak near zero energy. The theory [7] is much based on the assumption of the repulsive electron interaction in the normal metal. The question whether the reentrant effect is due to specific properties of noble metals or is shared by any normal metal experiencing the proximity effect can be answered only experimentally. We can just note that the theories in [6, 7] do not describe the temperature and field dependences of the paramagnetic susceptibility of the *NS* structure and do not explain the origin of the anomalously large amplitude of the reentrant effect.

It is worth mentioning the assumption made by Maki and Haas [8] that below the transition temperature (~ 10 *mK*) some noble metals (*Cu*, *Ag*, *Au*) can exhibit *p*-wave superconducting ordering, which may be responsible for the reentrant effect. This theory does not explain the high paramagnetic reentrant effect either.

In this Letter a reentrant effect theory is proposed, which is essentially based on the properties of the quantized levels of the *NS* structure. The levels with energies no more than $\Delta_0$ ($2\Delta_0$ is the gap of the superconductor) appear inside the normal metal bounded by the dielectric (vacuum) on one side and contacting the superconductor on the other side. The number of levels $n_0$ in the well is finite. Because of the Aharonov-Bohm effect [9], the spectrum of the *NS* structure is a function of the magnetic flux in a weak field. The specific feature of the quantum levels of the structure is that in a varying field *H* (or temperature) each level in the well periodically comes into coincidence with the chemical potential of the metal. As a result, the state of the system suffers strong degeneracy and the density of states of the *NS* sample experiences resonance spikes. We attribute the reentrant effect to this resonance.

The resonance spikes of the density of states as a function of the magnetic field were predicted earlier in our study on a structure consisting of a normal metal cylinder coated with a thin superconducting layer [10].

Let us consider a superconducting cylinder of radius *R* coated with a thin layer *d* of a pure normal metal. The structure is placed in a weak magnetic field $\vec{H}(0,0,H)$ applied along its symmetry axis. We proceed from a simplified *NS* model assuming that the order parameter modulus of the superconductor changes in a jump at the *NS* boundary. We introduce an angle of incidence of a quasi-particle $\alpha$ onto the dielectric boundary counted off the cylinder radius. It is evident that there are two classes of quasiparticle paths inside the normal metal. One class includes those in which $\alpha$ varies within $0 \lesssim \alpha \lesssim \alpha_{cr}$ ($\alpha_{cr}$ is the angle at which the path touches the *NS* boundary). In this case the quasiparticle collides successively with the dielectric and the *NS* boundary. The other class with $\alpha > \alpha_{cr}$ consists of the paths whose spectra are formed only by the quasiparticle-dielectric collisions. The spectra of the two classes paths differ considerably.

First we consider the paths with $\alpha \lesssim \alpha_{cr}$. The spectrum of the quasiparticles of the *NS* structure is readily obtainable by the method of multidimensional quasi-classics [10, 11] generalized for the presence of the Andreev scattering in the system [12]. We have

$$\varepsilon_n(q,\alpha) = \frac{\pi \hbar V_\mathcal{L}(q) \cos\alpha}{2d}\left(n + \frac{1}{\pi}\arccos\varepsilon/\Delta_0 - \frac{\Phi(\alpha)}{\Phi_0}\right). \qquad (1)$$

Here $V_\mathcal{L}(q) = \sqrt{p_F^2 - q^2}/m^*$, $p_F$ is the Fermi momentum, $q$ is the momentum component along the cylinder axis, $m^*$ is effective mass of the quasiparticle, $\Phi_0$ is the superconducting flux quantum. The magnetic flux through the area bounded by a part of the *NS* boundary and by the quasiparticle path at an angle $\alpha$ is given by $\Phi(\alpha) = 2\,\text{tg}\,\alpha \int_0^d A(x)dx$. The integral of $A(x)$ can be calculated if we know the distribution of the vector potential field inside the normal metal. The problem of the Meissner effect in superconductor-normal metal (proximity) sandwiches was solved by Zaikin [13]. The screening current in the *NS* structure was calculated in terms of the microscopic theory, and the expression $A(x) = Hx + \frac{4\pi}{c} j(a)\, x\,(d - x/2)$ can be obtained from the Maxwell equation $\text{rot}\,\vec{H} = \frac{4\pi}{c}\vec{j}$ with the boundary conditions $A(x=0) = 0$ and $\partial_x A(x=d) = H$. The screening current is a function of $a = \int_0^d A(x)dx$, $j(a) = -j_s \varphi(a/\Phi_0)$, where $j_s$ is superfluid current and $\varphi(x)$ is oscillating function of flux (at $a/\Phi_0 \ll 1$ we have $j(a) = -j_s a/\Phi_0$). Thus, we can write down the self-consistent equation for $a$ [13 – 15]

$$a = H\frac{d^2}{2} + \frac{4\pi}{3c} j(a)\, d^3 \qquad (2)$$

$j(a)$ turns to zero recurrently when $a/\Phi_0 = \frac{1}{2}, 1, \frac{3}{2}, 2, \ldots$ The spectrum of Eq.(1) is similar to Kulik's spectrum [16] for the current state of *SNS* contact. However, Eq. (1) includes an angle-dependent magnetic flux instead of the phase difference of the contacting superconductors.

We proceed from the expression for the thermodynamic potential $\Omega = -T\sum_S \ln\left[1 + \exp\left[-\varepsilon_S/T\right]\right]$, summation is made with respect to all quantum states of the spectrum, Eq. (1), and the spin (the Boltzmann constant $k_B = 1$). We use the approximation of equidistant levels in which the second term in Eq. (1) can be replaced with $1/2$. The contribution to susceptibility (per unit volume $V$ of the normal metal) can be found as $\chi = -\frac{1}{V}\frac{\partial^2 \Omega}{\partial H^2}$. Taking into account two orientations of the spin and two possible signs of $\alpha$ and $q$, as well as the finite number of levels $n_0$, we have the starting expression for susceptibility ($L$ is the cylinder height, $\zeta$ is the chemical potential of the metal):

$$\chi \sim \sum_{n=-n_0}^{+n_0} \int_{-\zeta}^{+\infty} \frac{d\varepsilon}{\text{ch}^2 \frac{\varepsilon}{2T}} \int_0^{\alpha_{cr}} d\alpha \sin^2 \alpha \int_0^{p_F} dq V_{\mathcal{L}}^2(q) \delta(\varepsilon - \varepsilon_n(q,\alpha)). \tag{3}$$

We take an integral over $q$ using the $\delta$-function and introduce the dimensionless energy $\epsilon = \varepsilon / \delta\varepsilon$ ($\delta\varepsilon = \pi\hbar V_F / 2d$). Since $\zeta / \delta\varepsilon \gg 1$, the lower limit of the integral of energy can be replaced with $-\infty$. By introducing the variable $x \equiv \text{tg}\,\alpha$ and the notations $a_n = n + 1/2$, $b = b(H, T) = a/\Phi_0$, $x_0 = \text{tg}\,\alpha_{cr} \simeq \sqrt{2R/d}$ and taking into account the parity of the integrand instead of Eq. (3) we obtain:

$$\chi = A \sum_{n=0,1,\ldots}^{n_0} \int_0^{+\infty} \frac{d\epsilon |\epsilon|^3}{\text{ch}^2[\eta \epsilon / 2]} \int_0^{x_0} \frac{dx\, x^2 \theta\left[|a_n - bx| - |\epsilon|\sqrt{1+x^2}\right]}{|a_n - bx|^3 \sqrt{|a_n - bx|^2 - \epsilon^2(1+x^2)}}. \tag{4}$$

Here $A = \frac{2\zeta d^2}{\pi R \Phi_0^2} \eta$, $\eta = \delta\varepsilon / T$, $\theta$ is the Heaviside step function. It is seen in Eq. (4) that for a given "subzone" $n$ the amplitude of the paramagnetic susceptibility increases sharply whenever the Andreev level coincides with the chemical potential of the metal. The resonant spike of susceptibility occurs when $a_n - bx$ tends to zero on a change in the magnetic field (or temperature). Because of the finite number of Andreev levels, the region of the existence of the isothermal reentrant effect, is within $0 < H \lesssim H_{n_0}$.

The undefined integral over $x$ can be calculated exactly. The roots of the square trinomial under the radical to within the first-order terms in $\epsilon$ are $x_{1,2} \simeq \alpha_0 \pm |\epsilon|\sqrt{1+\alpha_0^2}/b$ ($\alpha_0 = a_n/b$; $0 < x_1 < \alpha_0 < x_2 < x_0$). The $\theta$-function bars the region $x_1 < x < x_2$ from consideration. On substituting the limits of integration, the obtained expressions have different powers of the parameter $|\epsilon|^{-1}$. We retain only the most important terms of the order of $|\epsilon|^{-3}$. The integral over energy is taken easily when these terms are substituted in it. Finally, the susceptibility of the NS structure becomes

$$\chi \cong \frac{\zeta d^2}{\pi R \Phi_0^2} \sum_{n=0,1,\ldots}^{n_0} \frac{(n+1/2)^2 \,\text{th}\left[\frac{\pi\hbar V_F}{4dT}(n+1/2)\right]}{\left[(n+1/2)^2 + b^2(H,T)\right]^{3/2}}. \tag{5}$$

The flux $b = a/\Phi_0$ (Eq. (2)) depends on both the magnetic field and temperature. The screening current of the NS structure is $j = -j_s \varphi(a/\Phi_0)$, where $j_s \sim T^{-1} \exp(-4\pi T d/\hbar V_F)$ for $T \gg \hbar V_F / d$ [13,15]. It is seen from Eq. (5) that $b(H,T)$ increases with rising temperature ($H$ is fixed). At the same time the hyperbolic tangent argument becomes smaller than unity starting with a certain temperature. As a result, the amplitude of paramagnetic susceptibility decreases rapidly as the temperature rises. For the isothermal reentrant effect appears, the decreasing magnetic field causes an increase in the amplitude of paramagnetic susceptibility. Choosing the characteristic parameters of the problem $d=3.3\cdot10^{-4}$ cm, $R=8.2\cdot10^{-4}$ cm, $\zeta \sim 10^{-12}$ erg, $\Phi_0=2\cdot10^{-7}$ Gauss/cm$^2$, we can estimate the coefficient preceding the sum in Eq. (5) to be $\simeq 10^{-2}$. The sum itself has the logarithmic scale $\sim \ln n_0$. The effect is caused by the paths of the quasi-particles with $\alpha \lesssim \alpha_{cr}$.

The paths of the quasiparticles with $\alpha > \alpha_{cr}$ do not collide with the *NS* boundary and their quantum states are essentially similar to the states of the "whispering gallery" type that appear in the cross-section of a normal solid cylinder in a weak magnetic field [17, 18]. The caustic of this paths is approximately equal to the cylinder radius and the spectrum of the states carries no information about the parameters of the superconductor. The energy levels cannot be made coincident with the chemical potential of the metal by varying the magnetic field or temperature. As a result, the paramagnetic contribution into the thermodynamics of the paths with $\alpha > \alpha_{cr}$ has a much smaller amplitude.

The author is indebted to A.N. Omelyanchouk and A.A. Slutskin for helpful discussions.